\documentclass[pdflatex,sn-mathphys]{sn-jnl}% Math and Physical Sciences Reference Style
\usepackage{amsmath}
\usepackage{array}

\jyear{2021}%

\theoremstyle{thmstyleone}%
\theoremstyle{thmstyletwo}%

\theoremstyle{thmstylethree}%

\begin{document}

\title[How synchronized human networks escape local minima]{How synchronized human networks escape local minima}

\author[1]{\fnm{Elad} \sur{Schniderman}}

\author[2]{\fnm{Yahav} \sur{Avraham}}

\author[2]{\fnm{Shir} \sur{Shahal}}

\author[2]{\fnm{Hamootal} \sur{Duadi}}

\author[3]{\fnm{Nir} \sur{Davidson}}

\author[2]{\fnm{Moti} \sur{Fridman}}\email{mordechai.fridman@biu.ac.il}

\affil[1]{\orgdiv{Departments of Humanities and Arts}, \orgname{Technion – Israel Institute of Technology}, \orgaddress{\city{Haifa}, \postcode{32000}, \country{Israel}}}

\affil[2]{\orgdiv{Faculty of Engineering and the Institute of Nanotechnology and Advanced Materials}, \orgname{Bar Ilan University}, \orgaddress{\city{Ramat Gan}, \postcode{5290002}, \country{Israel}}}

\affil[3]{\orgdiv{Department of Physics of Complex Systems}, \orgname{Weizmann Institute of Science}, \orgaddress{\city{Rehovot}, \postcode{76100}, \country{Israel}}}


\abstract{Finding the global minimum in complex networks while avoiding local minima is challenging in many types of networks. We study the dynamics of complex human networks and observed that humans have different methods to avoid local minima than other networks. Humans can change the coupling strength between them or change their tempo. This leads to different dynamics than other networks and makes human networks more robust and better resilient against perturbations. We observed high-order vortex states, oscillation death, and amplitude death, due to the unique dynamics of the network. This research may have implications in politics, economics, pandemic control, decision-making, and predicting the dynamics of networks with artificial intelligence.}




\maketitle

\section{Introduction}\label{sec1}

Human interactions form complex networks of connections between the members of the networks. Synchronizing human networks can happen spontaneously~\cite{neda2000self_clap1,strogatz2005theoretical_MB} or intentionally~\cite{javarone2017evolutionary_env0,werner2007dynamics_env}. The synchronization of the network is essential for coordinating ideas~\cite{sumpter2012six_dm0,smaldino2012origins_dm1}, and the well-being of its members~\cite{wasserman1994social_HS1,morris2005social_HS2,solferino2021human_toxic}, and is seen in other organisms as well~\cite{sumpter2006principles_bio1,sarfati2022chimera_bio_ff,davis2001biological_cs}. Understanding the dynamics of human networks has implications for politics, economics, social sciences, and pandemic control. The topology of human networks determines if the network will be able to synchronize~\cite{dorfler2014synchronization}, who is most probable to become a leader~\cite{calabrese2021spontaneous_leader}, which decision the network is more likely to make~\cite{sumpter2012six_dm0,conradt2005consensus_dm2}, and how stable is the network to small perturbations~\cite{gambuzza2021stability_stable1,sorrentino2010stability_stable2,kassabov2022global_stable3}. 

The synchronization dynamics of networks can be analyzed in terms of an effective potential landscape in which the system evolves~\cite{roy1992dynamical,fridman2010passive}. To reach the fully synchronized state, which is the global minimum of this potential landscape, a human network must avoid getting trapped in local minima, where not all humans are synchronized ~\cite{ling2019landscape_energy2,ravoori2011robustness_energy3,leylaz2021identification_energy1,roy1992dynamical,fridman2010passive,fridman2012measuring}. Avoiding such local minima also has implications for increasing the stability in other types of networks~\cite{jalan2005synchronized_other}, speeding up machine learning processes~\cite{ayodele2010types_AI}, and optimization problems in spin-glass dynamics~\cite{heim2015quantum_spinglass}.

Human networks can reach synchronization by finding unique solutions which are more stable compared to other networks due to the human ability to focus on some inputs while ignoring
others~\cite{shahal2020synchronization_prev}. However, it is still unclear how human networks reach these stable solutions, what the network dynamics that lead to it are, and in particular, how the human network escapes local minima.  
 
In this paper, we study the synchronization dynamics of human networks with closed-ring topology and unidirectional time-delayed coupling (Fig.~\ref{fig:All6}(a)). Such a network has a complex potential landscape with well-defined local and global minima~\cite{tradonsky2015conversion,pal2015phase,kassabov2022global_stable3} that we can tune and control in real time. We prepare the system in a global minimum (fully synchronized) state and then adiabatically transform the potential landscape by tuning the coupling delay time such that this state becomes a local minimum. We then study in detail how the system escapes this local minimum into the new fully synchronized global minimum by measuring the amplitude, tempo, and phase of each node and identifying which node is following which. 
 
Studying the rhythmic behavior of humans in general, and music in particular, can reveal aspects of human network dynamics which are usually hard to identify~\cite{d2015can_mus1}. We show that humans can escape local minima by self-tuning local properties of the networks such as their tempo, amplitude, and the coupling strength between players. The ability to tune these parameters changes dramatically the dynamics of the network and has implications for other networks where each node has decision-making abilities~\cite{kolling2015human_robot}. 

\section{Experiment} \label{sec::experiment}

\begin{figure}[htb]
    \centering
    \includegraphics[width=0.6\linewidth]{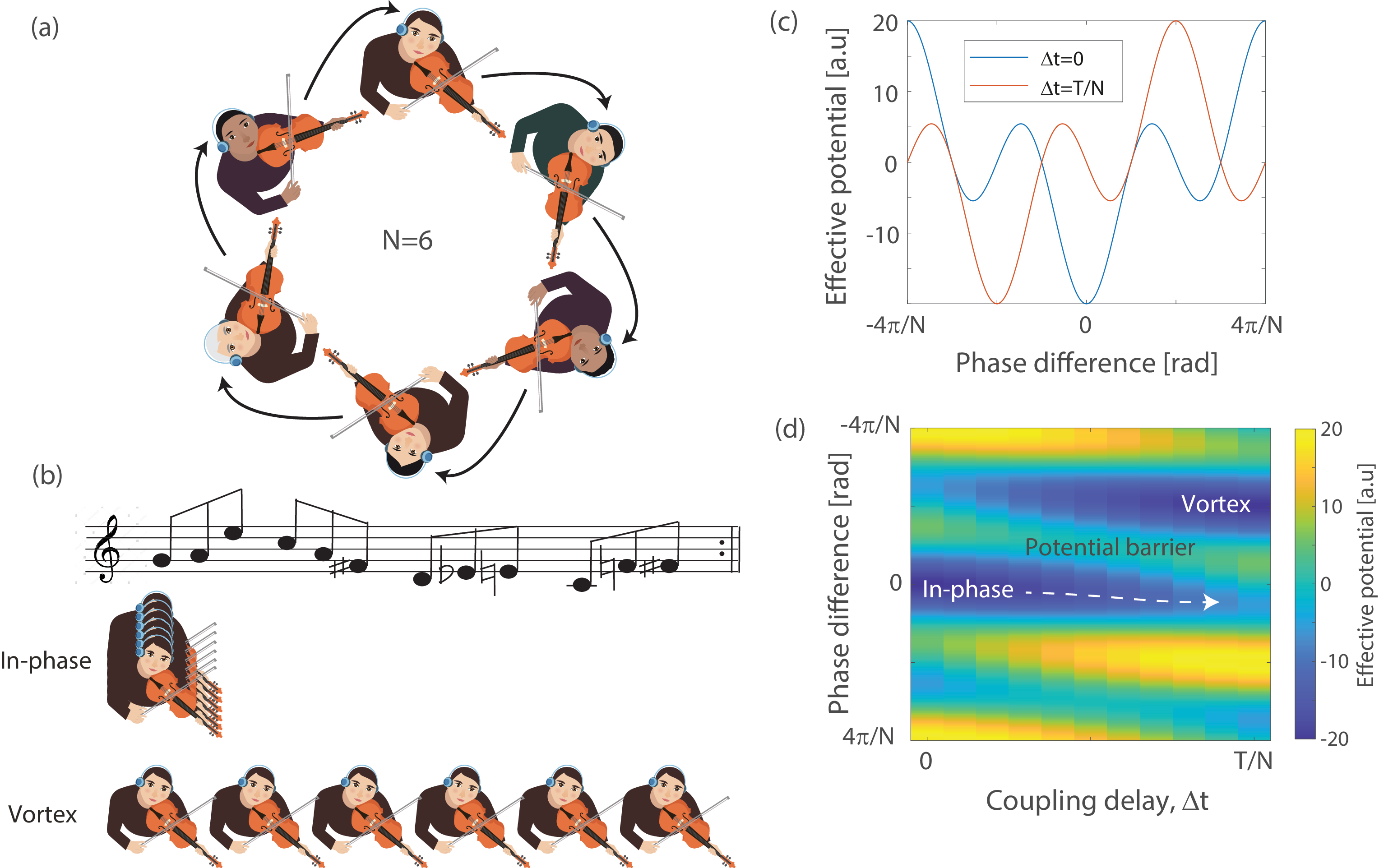}
    \caption{Coupled violin players on a unidirectional ring network, where each player is hearing only the player on its right with a controllable delayed coupling. (a) Schematic of $N=6$ unidirectional coupled players. (b) The musical phrase the players are periodically repeating illustrates the two possible states of synchronization: the in-phase state where all the players are playing the same note at the same time, and the vortex state, where due to the delay between the players, each player is playing a different note. (c) The effective potential (defined in section~\ref{sec::spreading}) of the system as a function of the phase difference between the players for two values of delay $\Delta t$. For $\Delta t=0$, the minimum potential is at zero phase difference, indicating an in-phase state of synchronization. For $\Delta t=T/N$, the minimum potential is at $-2\pi/N$ indicating a vortex state and the in-phase state becomes a local minimum, protected by a potential barrier. (d) Effective potential map as a function of the phase difference and delayed coupling between the players.}
    \label{fig:All6}
\end{figure}

Our system with coupled violin players is schematically shown in Fig.~\ref{fig:All6}(a)~\cite{shahal2020synchronization_prev}. The violin players cannot see or hear the others apart from the audio signal they receive through their headphones. The sound of each player is connected to a control system that receives the audio signal from all players and distributes it to each player. The control system provides a tunable, programmable, and accurate real-time control of the network connectivity (who is connected to who) and the strength and delay of the coupling between the players. The players repeat the music phrase shown in Fig.~\ref{fig:All6}(b), and are instructed to synchronize with what they hear. In this study, we set the players' network as a unidirectional ring, so each player hears only a single neighbor with periodic boundary conditions, illustrated in Fig~\ref{fig:All6}(a) for $N=6$ players. We performed measurements also with different numbers of players including $N=2,3,4,5,6,7,8$, and 16. We measure the note each player is playing as a function of time, and accordingly, obtain the phase $\varphi(t)$ of the player, where $\varphi=0$ denotes that the player is playing the beginning of the musical phrase, and $\varphi=2\pi$ denotes that the player is at the end of the musical phrase of duration $T$ ~\cite{shahal2020synchronization_prev}. 

Each experiment starts with zero time-delay $\Delta t=0$ between all players with coupling strength sufficiently high~\cite{shahal2020synchronization_prev} to ensure that all players are synchronized in-phase (i.e. playing the same note at the same time, see Fig~\ref{fig:All6}(b)). Then, we increase $\Delta t$ linearly with time as $\Delta t=t/30$. When $\Delta t$ exceeds $nT/N$  the in-phase state becomes a local minimum and the global minimum solution is a vortex state, where each player has identical delay compared to its neighbor, satisfying the periodic boundary conditions (see Fig~\ref{fig:All6}(b)). The $n$ parameter is an integer number and denotes the topological charge of the system~\cite{pal2015phase}. The players must then leave their in-phase state and find the vortex state.   

The dynamics of the coupled violin players performing a periodic rhythm can be analyzed by the Kuramoto model~\cite{kuramoto2012chemical_kur1,kuramoto1987statistical_kur0,strogatz2000kuramoto_kur2} which describes an over-damped motion of coupled phase oscillators in an effective potential~\cite{leylaz2021identification_energy1,roy1992dynamical,fridman2010passive,fridman2012measuring}. The derivation of the Kuramoto effective potential is presented below in section~\ref{sec::spreading}. Figure ~\ref{fig:All6}(c) depicts two representative effective potentials, for $\Delta t=0$ and  $\Delta t=T/N$, respectively, as a function of the phase difference between coupled players. For $\Delta t=0$, the global minimum potential is at zero phase difference between coupled players, corresponding to the in-phase synchronization state. For $\Delta t=T/N$, the global minimum of the potential is at a phase difference of $-2\pi/N$, corresponding to a vortex state of synchronization, while the in-phase state becomes a local minimum. 

Figure ~\ref{fig:All6}(d) presents the effective potential as a function of $\Delta t$. The dashed line denotes the system trajectory in the phase space as the delay increases adiabatically from $\Delta t=0$. The system starts from the in-phase state and in order for the system to reach the emerging global minimum vortex state, it must overcome a potential barrier. Here, we present and analyze four types of dynamics in human networks for overcoming this potential barrier and escaping the local minimum into the global one. 

In the first dynamics, some of the players ignore the signals they receive, thereby reducing the effective coupling strength to their neighbor. Then, they can freely spread their phase to reach the vortex state (see section~\ref{sec::spreading}). In the second, the players are slowing down their tempo so the in-phase state remains the global minimum for arbitrarily large delays (see section~\ref{sec::slow}). In the third, the players further slow down until everyone plays the same note indefinitely. This state, known as oscillation death~\cite{ermentrout1990oscillator_death}, is a stable synchronized solution regardless of the delay between the players. The players then spontaneously emerge from the oscillation death directly into the globally stable vortex state (see section~\ref{sec::death}). In the fourth, some of the players stop playing (i.e. reduce their amplitude to near zero). In this state, known as amplitude death, the network topology changes into an open ring, where the local minima and its potential barrier disappear. Thus, enabling the other players to find the vortex state (see section~\ref{sec::death}). 

We observed these four distinct strategies to escape local minima at all network sizes from $N=2$ up to $N=16$ players. While representing different emerging strategies to overcome barriers and reach stable minima, they all rely on the unique ability of human players to adjust their playing amplitude and tempo and change the effective coupling strength between them by ignoring frustrating inputs. More results are shown in the supplemental materials.

\section{Spreading the phase} \label{sec::spreading}

The first dynamic we observe is when the players are spreading their phases to escape from the in-phase local minimum into a vortex-phase state global minimum. This dynamic is enabled by the players' ability to reduce the coupling between them until they find a stable state. 

\begin{figure}[htb]
    \centering
    \includegraphics[width=0.8\linewidth]{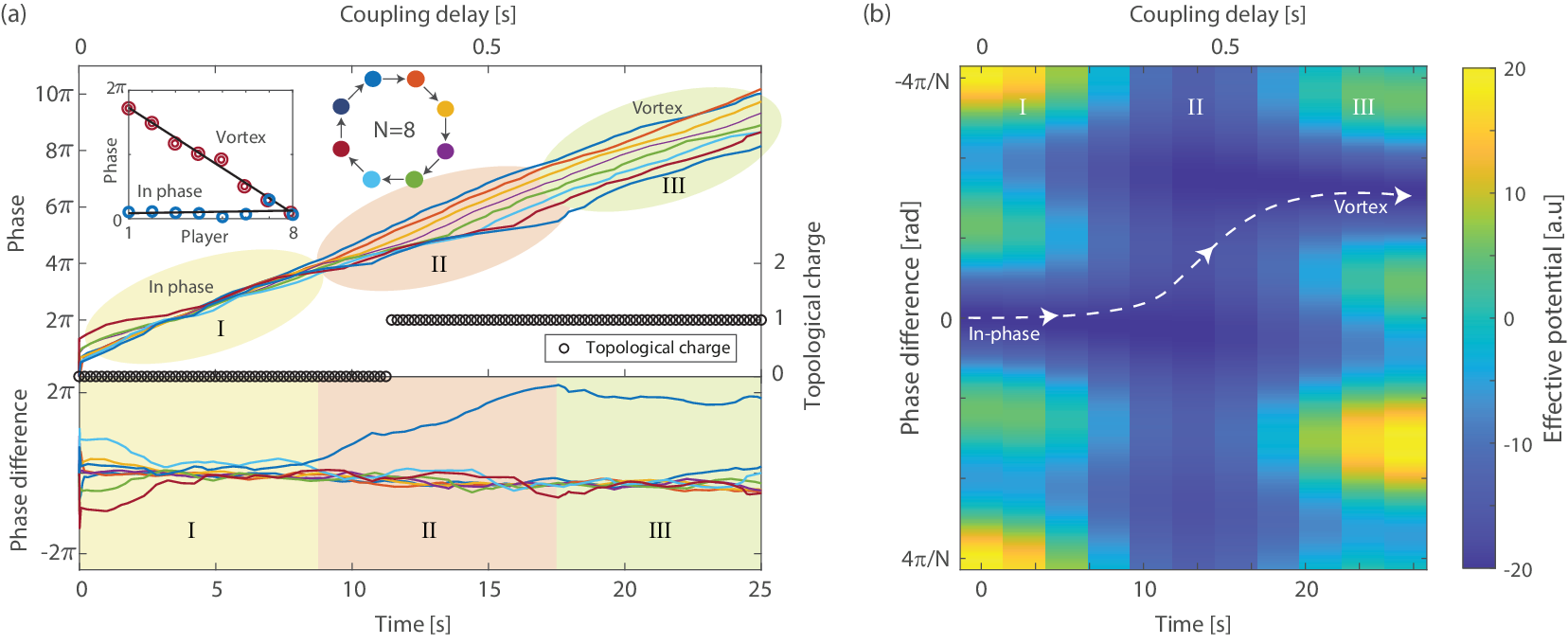}
    \caption{Coupled $N=6$ players situated on a ring in a unidirectional coupling, according to the inset scheme, starting from an in-phase state of synchronization and finding the vortex state. (a) (top) Measured phase of each player as a function of time as we increase the coupling delay between the players. Each color denotes the phase of a player according to the inset scheme. Black circles denote the topological charge of the system as a function of time. (bottom) The difference between the phase of each player and the phase of its delayed neighbor. (inset) The phase of all players in a state of in-phase synchronization (stage I, blue circles) and the phase of all players in a vortex state (stage III, red circles). (b) The effective potential of the system as a function of time and coupling delay. In stage II the effective coupling strength is reduced (see text) thus eliminating the potential barrier between the in-phase and the vortex states to enable the system to reach the global minimum vortex state. The trajectory of the system is denoted by the white dashed curve.}
    \label{fig:all8}
\end{figure}

Figure~\ref{fig:all8}(a)(top) presents the measured phase of $N=8$ players situated on a ring with unidirectional coupling as a function of time $t$ and delay $\Delta t=t/30$. Fig.~\ref{fig:all8}(a)(bottom) presents the measured phase difference between each player and its delayed neighbor. As seen, during the first 3 seconds all phase difference converges to near zero, indicating that all players are synchronized in phase, and remain so until $t=8$. This first stage is marked as stage I. The average phase of each player during this stage, presented as the blue circles in the inset of Fig.~\ref{fig:all8}(a), confirms that all the players have nearly the same phase. When the delay is further increased, one of the players starts to ignore its neighbor, indicated by the linearly increasing blue curve in the phase difference in Fig.~\ref{fig:all8}(a)(bottom). Thus, the other players can freely change their phase to follow the increasing delay between them. This is marked as stage II. The phase of the players spread until reaching a total phase difference of $2\pi$ between the first and the last player, thus forming a stable vortex solution with a topological charge $n=1$~\cite{pal2015phase}, shown as the black circles in Fig.~\ref{fig:all8}(a). This vortex solution remains stable as the delay is further increased, indicated by a constant phase difference between all coupled neighbors (marked as stage III). The average phase of all players in stage III is shown as red circles in the inset of Fig.~\ref{fig:all8}(a), and verifies the expected linear phase of the vortex state. 

%Finally, the topological charge of the system indicates the number of $2\pi$ in which the players are spread. This topological charge equals zero when the players are in-phase, and unity when they reach the vortex state and is shown as the black circles in Fig.~\ref{fig:all8}(a).    

Following~\cite{leylaz2021identification_energy1,roy1992dynamical,fridman2010passive,fridman2012measuring}, we use the Kuramoto model for $N$ oscillators on a ring with unidirectional time-delayed coupling ~\cite{kuramoto2012chemical_kur1,kuramoto1987statistical_kur0,strogatz2000kuramoto_kur2} and uniform frequency $\omega$ to introduce an effective potential whose local and global minima dictate the players' dynamics. The phase of each oscillator $\varphi_n(t)$ follows:
\begin{equation}
\frac{\partial\varphi_n(t)}{\partial t}=\omega+\kappa \sin (\varphi_{n+1}(t-\Delta t)-\varphi_n(t)),
\label{eq:phase_time}
\end{equation}
with $\kappa$, and $\Delta t$ denoting the strength and delay of the coupling. The periodic boundary conditions, $\varphi_{N+1}=\varphi_1$ dictate $ \sum_{n=1}^N \Delta\varphi_n=0$, where $ \Delta\varphi_n(t)=\varphi_{n+1}(t)-\varphi_n(t)$.  
Assuming uniformity $\Delta \varphi_n=\Delta \varphi$, without loss of generality, we obtain: 
\begin{equation}
\frac{\partial\Delta\varphi(t)}{\partial t}=-\frac{\partial V(\Delta\varphi)}{\partial \Delta \varphi}. 
\end{equation}
where:
\begin{equation}
V(\Delta\varphi)=-\Omega_n \Delta\varphi-\frac{\kappa}{N-1} \cos((N-1)\Delta\varphi+\omega \Delta t)+\kappa\cos(\Delta\varphi+\omega \Delta t),
\label{eq:Pot}
\end{equation}
is the effective potential governing the dynamics of the system. The detailed analytical derivation is presented in the supplemental materials.

The calculated effective potential for constant tempo is shown in Fig.~\ref{fig:All6}(d) as a function of $\Delta t$ and $\Delta\varphi$. For $\Delta t=0$ it reveals a global minimum at $\Delta\varphi=0$, where all oscillators have the same frequency and phase, namely, an in-phase state of synchronization. However, when the delay increases beyond $\omega \Delta t > \pi/(2N)$, the $\Delta\varphi=0$ becomes a local minimum and the vortex state emerges as a new global minimum at $\Delta\varphi=-2\pi/N$. Between the in-phase state and the vortex state, there is a potential barrier. Therefore, increasing the delay adiabatically, while the system stays in the in-phase state of synchronization, transfers the system to a local minimum.   

To account for the players' dynamics observed in Fig.~\ref{fig:all8}, we assume that the coupling strength in stage II is reduced by the player's tendency to ignore frustrating inputs. Specifically, we assume $\kappa(t)=\kappa(t=0)\cos^2(\Delta t N \pi/T)$. The resulting modified effective potential is shown in Fig.~\ref{fig:all8}(b). The dashed curve follows the system trajectory in the phase space. The system starts in an in-phase state of synchronization and as the delay increases, the coupling drops leading to a lower potential barrier. Thus, the system can adiabatically evolve into the vortex state. Finally, the coupling increases back to its original value $\kappa(t=0)$.  

\begin{figure}[htb]
    \centering
    \includegraphics[width=0.8\linewidth]{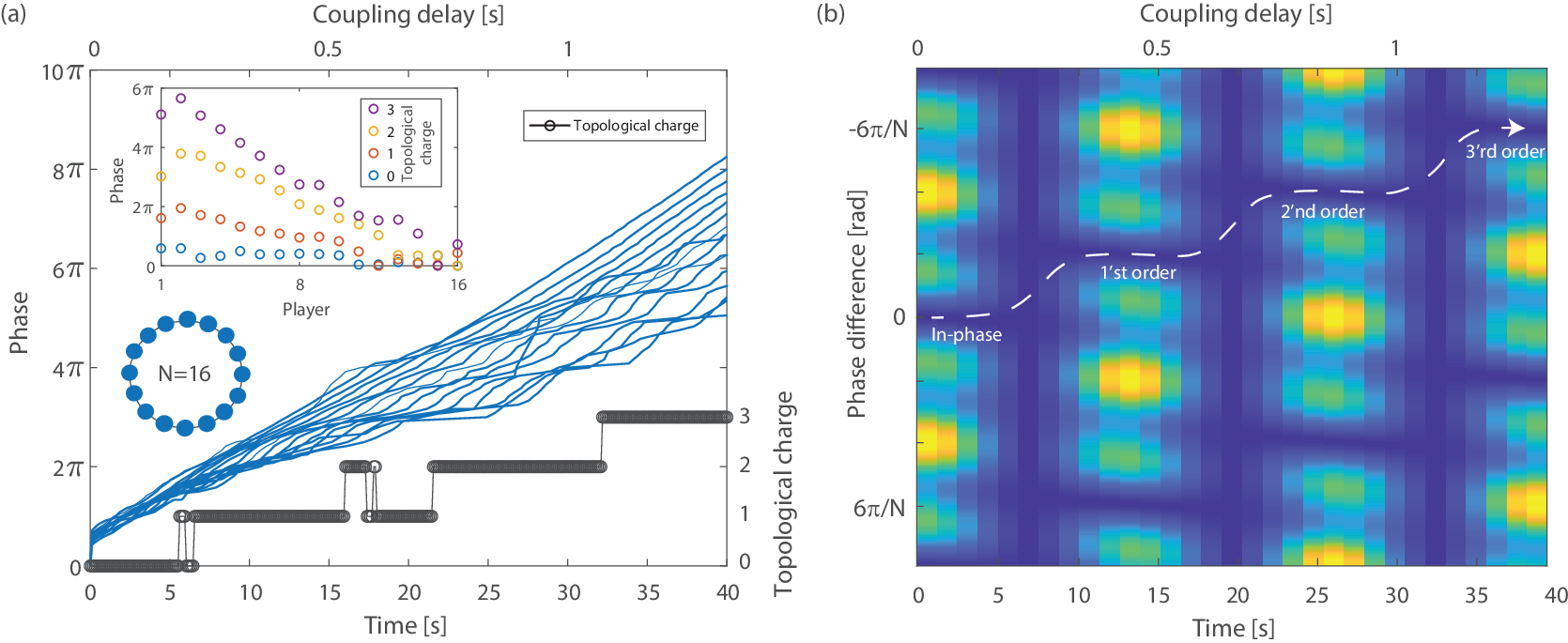}
    \caption{Coupled $N=16$ players situated on a ring in a unidirectional coupling, according to the inset scheme, showing spreading of the phase and reaching high order vortex states. (a) The measured phase of $N=16$ coupled violin players together with the topological charge of the network as we increase the delayed coupling between them, showing higher-order vortex states. (Inset) the phase of the 16 players at $t=1.5, 13.5, 24$, and 34 seconds, with topological charges $n=$ 0, 1, 2, and 3, respectively. (b) The effective potential of the system as a function of time and coupling delay shows how the system evolves.}
    \label{fig:highOrder}
\end{figure}

When the delay between the players is further increased, the first-order vortex becomes unstable and the next-order vortex becomes stable. We observed such multiple transitions to higher-order vortex states with $N=16$ coupled violin players on a unidirectional ring. The measured phases of all the players together with the topological charge of the network are shown in Fig.~\ref{fig:highOrder}(a). In the inset, we show the phase of the players at four representative times, 1.5, 13.5, 24, and 34 seconds, with topological charges  $n=$ 0, 1, 2, and 3, respectively. As the players' phases spread over a wider range, the network reaches a higher vortex state, denoted by a higher topological charge. We also calculate the effective potential as a function of time, again assuming $\kappa(t)=\cos^2(\Delta t N \pi/T)$, showing how the system can evolve from the in-phase to each order of vortex.  

\section{Slowing the tempo}  \label{sec::slow}

\begin{figure}[htb]
    \centering
    \includegraphics[width=\linewidth]{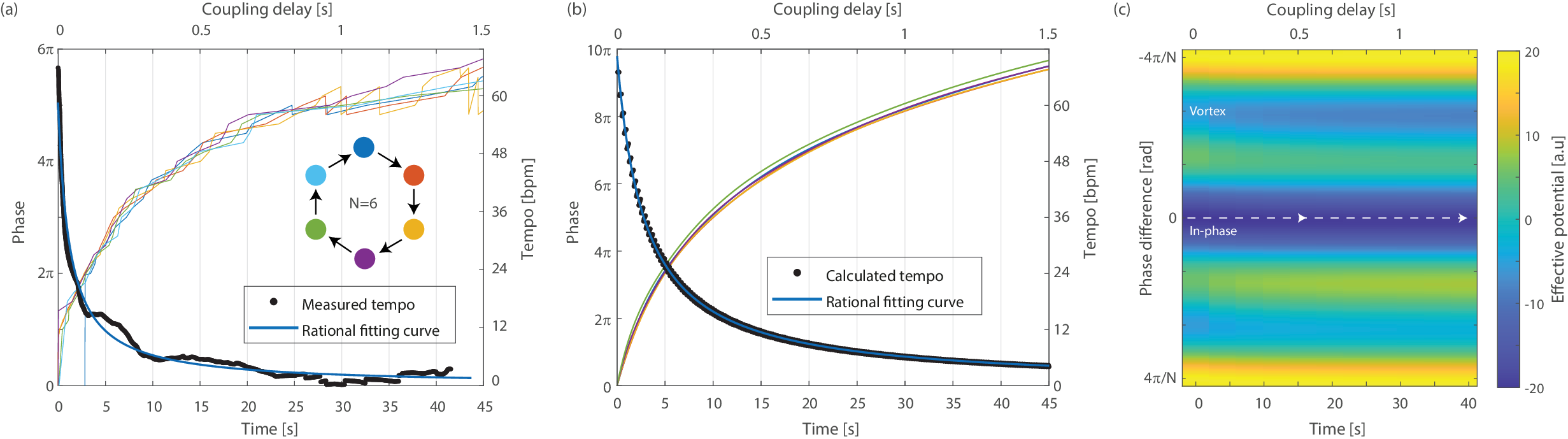}
    \caption{Coupled $N=6$ players situated on a ring in a unidirectional coupling showing tempo slowing down. (a) Experimental measurements of the phase and average tempo as we increase the delayed coupling between them. Each color denotes the phase of a player according to the inset scheme. (b) Numerical calculations of the system. (c) The effective potential of the system as a function of time when the tempo of the players is following Eq.~\ref{eq:slow} indicating that the in-phase state stays the global minima for arbitrary large delays.}
    \label{fig:Slow}
\end{figure}

The second dynamic we observe is the slowing down of the players' tempo as an alternative strategy to maintain a stable in-phase synchronization state in the presence of delayed coupling~\cite{chafe2010effect_delay}. The measured phase results as a function of time for six coupled players are shown in Fig.~\ref{fig:Slow}(a). From these results, we evaluate the tempo of the players by calculating the average derivative of the phase. As seen, the initial (natural) tempo of 60 bpm slows down significantly for long delay times reaching about 6 bpm for $\Delta t > 1s$. This slowing down lets players maintain the in-phase state of synchronization by keeping the coupling delay small relative to the tempo.   

To quantitatively analyze the tempo slowing, we resort again to the Kuramoto model~\cite{kuramoto2012chemical_kur1,kuramoto1987statistical_kur0,strogatz2000kuramoto_kur2}. Assuming $\Delta t<T/N$, we can expend $\varphi_n(t-\Delta t) \approx \varphi_n(t)-\Delta t \partial \varphi_n / \partial t$ and  $\sin(\alpha) \approx \alpha$ to obtain from Eq.~\ref{eq:phase_time} the phase difference as a function of time for $\Delta\varphi_{N-1}$:
\begin{equation}
\label{eq:phaseDiff}
    \frac{\partial\Delta \varphi(t)}{\partial t}=\frac{\Omega-N \kappa \Delta \varphi}{1-(N-1)\kappa \Delta t},
\end{equation}
where $\Omega$ is the average tempo difference between the players. Then we obtain,
\begin{equation}
    \frac{\partial\varphi(t)}{\partial t}=\frac{\omega}{1+N \kappa \Delta t },
    \label{eq:slow}
\end{equation}
where $\varphi=\sum\varphi_n/N$ is the average phase of all oscillators. Thus, the tempo of the coupled oscillators slows down as long as the players stay phase locked, ensuring that the condition $\Delta t<T/N$ is satisfied and indicating that our assumptions are valid. Using Eq.~\ref{eq:slow}, to fit the measured tempo (blue curve in Fig.~\ref{fig:Slow}(a)) yields excellent agreement, where the fit parameters $\omega=0.29 Hz$ and $\kappa=0.63$ are consistent with our system. 

Next, We perform numerical simulations of the Kuramoto model (Eq.~\ref{eq:phase_time}) for six coupled players with a coupling strength of $\kappa=0.6$. The calculated phase of all the players as a function of time/delay together with the average tempo are shown in Fig.~\ref{fig:Slow}(b). As evident, the exact numerical simulations agree with the measured experimental results as well as with the analytical approximation of Eq.~\ref{eq:slow} ((blue curve in Fig.~\ref{fig:Slow}(b)). 

Finally, We calculate the effective potential as a function of time with the tempo obtained from Eq.~\ref{eq:slow}. As evident, the in-phase state of synchronization remains the global minimum even though the coupling delay increases. Therefore, the system is following the dashed line and remains in the in-phase synchronization state for all coupling delays.  

\section{Oscillation death and amplitude death}  \label{sec::death}

In this section, we present two additional mechanisms that have been observed in other networks on coupled nonlinear oscillators: oscillation death ~\cite{ermentrout1990oscillator_death} and amplitude death ~\cite{zou2013generalizing_death2}, and show how both enable the human network to escape from a local minimum (in-phase synchronization) into a global one (vortex state).

\begin{figure}[htb]
    \centering
    \includegraphics[width=\linewidth]{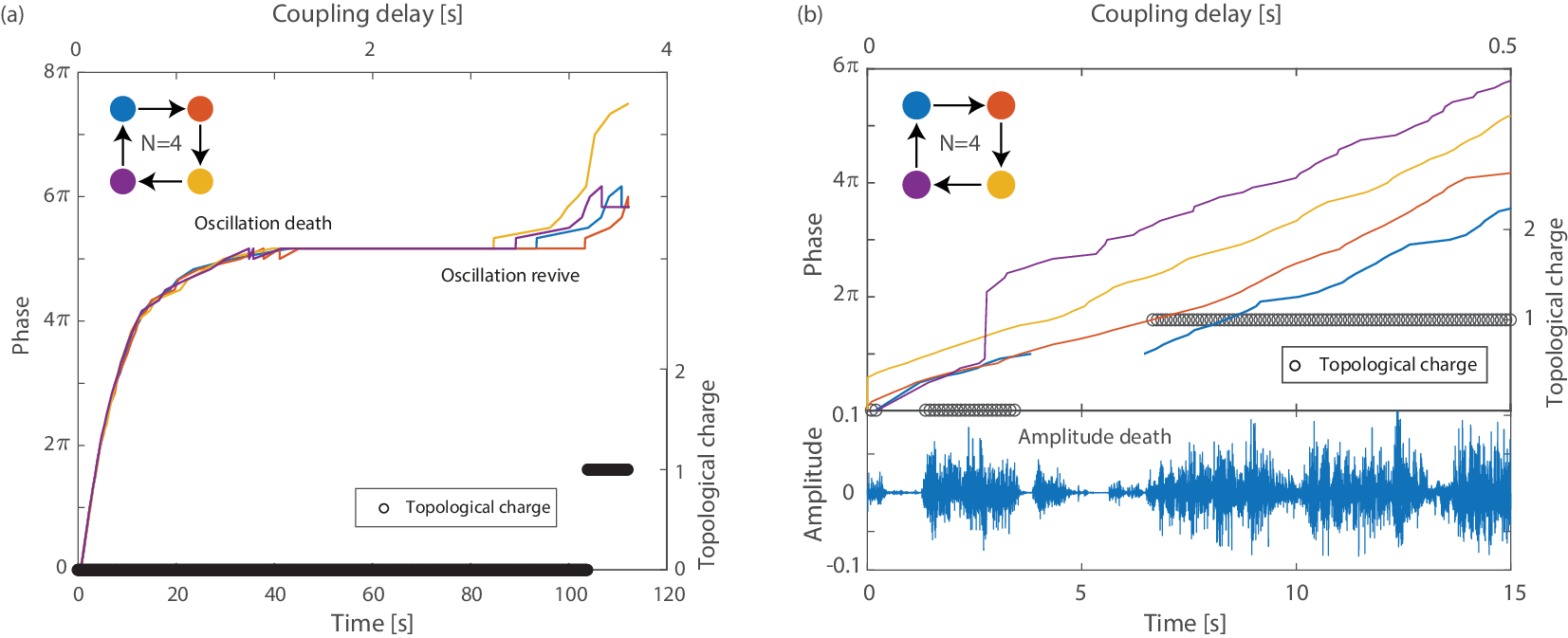}
    \caption{Coupled $N=4$ players situated on a ring in a unidirectional coupling showing oscillation and amplitude death. (a) Coupled violin players reach a state of oscillation death. The players slow down their tempo until all the players play the same note. After 20 seconds, the players revive the oscillation spontaneously into a vortex state. Each color denotes the phase of a player according to the inset scheme. (b) Coupled violin players show the amplitude death of one player. The upper graph shows the phase of all the players together with the topological charge as a function of time. The lower graph shows the playing amplitude of Player 1 as a function of time. We see that Player 1 stops playing around $t=5 s$ until the system finds the stable first-order vortex state.}
    \label{fig:AD}
\end{figure}

When the tempo slows too much, the players can get stuck in a state of oscillation death~\cite{ermentrout1990oscillator_death} where all the players are playing the same note indefinitely, thereby maintaining a degenerate form of synchronization. Representative results of such oscillation death for four coupled violin players are shown in Fig.~\ref{fig:AD}(a). The four players are slowing down their tempo, and after 40 seconds they all play the same note for 20 seconds. Then, the players spontaneously revive the oscillation when one of the players stopped following its neighbor. Reviving oscillations after oscillation death typically requires external perturbation to the system~\cite{zou2013generalizing_death2}, but here we demonstrate that human networks can revive the oscillations spontaneously. In addition, they revive the oscillations into the stable vortex state as indicated by the topological charge which jumps to $n=1$ when the oscillations revive.  

Violin players can adjust not just the phase and tempo but also their playing amplitude. In highly frustrating situations one or more of the players can reduce their amplitude significantly, known as amplitude death~\cite{zou2013generalizing_death2}. Representative measurements of such amplitude death are shown in Fig.~\ref{fig:AD}(b). Here, the coupling delay of four violin players is increased until one of the players, denoted as Player 1, cannot synchronize with Player 4 leading to frustration. Therefore, this player stops playing, as evidenced by its nearly vanishing amplitude, shown in the lower graph in Fig.~\ref{fig:AD}(b). When one of the players stops playing, the closed ring effectively changes into an open ring topology where all the other players are free to shift their phases according to the coupling delay. After a few seconds, they find the vortex state, as indicated by the measured topological charge, and Player 1 resumes playing.

\section{discussion and conclusions}

To conclude, we investigated the synchronization dynamics of coupled violin players with ring configuration and unidirectional time-delayed coupling. This configuration is governed by a potential landscape that includes well-defined local and global minima that are controlled by the coupling delay time. By starting with a zero coupling delay, we prepared the system in a globally stable in-phase synchronization state and then adiabatically increased the coupling delay such that the in-phase state became a local minimum. We observed four different routes for the system to escape this local minimum into the global one.   

In the first, the players change their coupling strength and shift from the local minimum of the in-phase state into the stable vortex state. In the second, the player slow down their tempo so the in-phase state remained the global minimum. In the third, all the players play the same note indefinitely which is a stable state regardless of the coupling delay. In the fourth, one of the players stopped playing, effectively changing the system topology into an open ring, thus allowing the players to find the vortex state, and then resume playing. 

Our results indicate that human networks are more robust than other networks since they have unique methods for escaping local minima into global ones. The results shed new light on the dynamics of human networks and how a group of humans can reach synchronization while escaping local minima. It has implications in decision-making theory, politics, economics, and for other networks where each node in the network can change its coupling strength or tempo.

\backmatter

\bmhead{Supplementary information}
All the raw measured data of all the experiments are available in 10.6084/m9.figshare.22822103. 

\bmhead{Acknowledgments}

%\section*{Declarations}
%
%Some journals require declarations to be submitted in a standardized format. Please check the Instructions for Authors of the journal to which you are submitting to see if you need to complete this section. If yes, your manuscript must contain the following sections under the heading `Declarations':

%\begin{itemize}
%\item Conflict of interest/Competing interests (check journal-specific guidelines for which heading to use)
%\item Ethics approval 
%\item Consent for publication
%\item Availability of data and materials
%\item Authors' contributions
%\end{itemize}

\noindent
If any of the sections are not relevant to your manuscript, please include the heading and write `Not applicable' for that section. 

\bigskip
\begin{flushleft}%
Editorial Policies for:

\bigskip\noindent
Springer journals and proceedings: \url{https://www.springer.com/gp/editorial-policies}

\bigskip\noindent
Nature Portfolio journals: \url{https://www.nature.com/nature-research/editorial-policies}

\bigskip\noindent
\textit{Scientific Reports}: \url{https://www.nature.com/srep/journal-policies/editorial-policies}

\bigskip\noindent
BMC journals: \url{https://www.biomedcentral.com/getpublished/editorial-policies}
\end{flushleft}

\begin{appendices}




\end{appendices}


\bibliography{sn-bibliography}% common bib file


\end{document}